# A SUPER-CONDUCTING LINAC DRIVER FOR THE HFBR[*]

J. Alessi, D. Raparia, A.G. Ruggiero, BNL, Po Box 5000, Upton, NY 11973, USA

*Abstract*

This paper reports on the feasibility study of a proton Super-Conducting Linac (SCL) as a driver for the High-Flux Breeder Reactor (HFBR) at Brookhaven National Laboratory (BNL). The Linac operates in Continuos Wave (CW) mode to produce an average 10 MW of beam power. The Linac beam energy is 1.0 GeV. The average proton beam intensity in exit is 10 mA.

## 1 INTRODUCTION

A proton SCL has been proposed to drive an internal solid target placed at the core of the de-commissioned HFBR at BNL. The purpose is to take advantage of the HFBR instrumentation and facility for the generation of a continuous, high-intensity neutron source.

The SCL driver accelerates protons to 1 GeV, operates in CW mode, and generates an average beam power of 10 MWatt. The average beam current is 10 mA, and the total length about 300 m. The Linac is made of three parts: a Front-End, that is a 15 mA ion source, followed by a 5-MeV RFQ, a room temperature 150-MeV Drift-Tube Linac (DTL), and the SCL proper. This is made of two sections: the Medium-Energy Section (MES) that accelerates protons to 300 MeV, and the High-Energy Section (HES) that accelerates to 1 GeV. The selected operating frequency is 805 MHz.

A spreadsheet program has been used for the design and cost estimate. The design has shown that the accelerator is feasible, can be built in a relatively short period of few years, and has an estimated total cost of about 300 M$.

## 2 REQUIREMENTS OF THE DRIVER

The accelerator driver of the HFBR facility is a proton SCL about 300 m long with a straight-line geometry. The proton beam aims directly to the core, that is the center of the HFBR facility. The actual location of the accelerator that does not interfere with other facilities and utilities remains to be investigated.

The SCL requirements are: 1-GeV Proton Energy, 10-MWatt Beam Power, and CW Mode of Operation. The Beam Current at the exit of the SCL is 10 mA. Acceleration of positive-ions (protons) is assumed.

A preliminary design has shown that the accelerator is feasible, can be build in a relatively short period of few years, and has a total cost of about 300-400 M$. A spread-sheet program [1] was used for the design and cost estimate, developed at the time of the design of the Accelerator for Tritium Production (APT) Linac [2, 4]. The requirements of the Linac driving the HFBR are similar to those of a Linac already investigated for a different type of application (Energy Amplifier) [3]. We have based our estimates essentially on that design.

The accelerator is made of four major parts, shown in Figure 1: the Front-End, a room-temperature Low-Energy Section, the SCL proper, and the Transport to the Target.

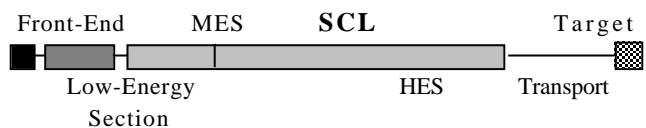

Figure 1. Layout of the 1-GeV, 10-MW SCL

The Front-End is made of an Ion Source placed on a platform at 35-50 kVolt. It has a continuous beam output of 15 mA. It is followed by a 201.25-MHZ RFQ which focus, bunch and accelerate the beam to about 3-5 MeV. At the exit, the beam bunches are compressed sufficiently to be squeezed within the rf buckets of the Low-Energy Section which operates at 402.5 or 805 MHz. Because of the relatively low beam current, and the absence of stringent requirements on the beam emittance and momentum spread, space-charge effects are not expected to play a relevant role. As a consequence, no major beam losses are expected in the RFQ. A transmission of 80% is conservatively assumed, and the beam intensity at the exit of the RFQ is 12 mA. We allow another 80% overall transmission efficiency, that is a 20% beam loss, during the transfer of the beam through the rest of the accelerator, all the way down to the Target. At the exit of the SCL, and on the Target, the beam intensity is then 10 mA.

## 3 LINAC DESIGN

In a proton Linac there is a large variation of beam velocity, in our case from $\beta = 0.08$ at 3 MeV to $\beta = 0.875$ at 1 GeV. The Low-Energy section cannot be made of half-wavelength super-conducting rf cavities, though quarter-wavelength super-conducting linear accelerators do exist and are successfully operational. We prefer to adopt here a room-temperature Low-Energy Section, made for instance of a conventional DTL similar to the

---

one operating as injector to the Alternating Gradient Synchrotron at BNL. We shall also adopt a 150 MeV energy for this section to ease the design and manufacturing of the rf cavities in the early part of the SCL proper. Other solutions are of course possible, and they should be examined in a more careful and detailed design. We shall not attempt here to say more about the Front-End and the Low-Energy Section, except noting that likely the cost of both together is in the range 50-100 M$ (year 2000).

Thus, the SCL proper begins at 150 MeV and ends at 1 GeV. The corresponding variation of velocity is from $\beta = 0.5066$ to $\beta = 0.8750$. Since the length of the rf cavity cells is $L = \beta\lambda/2$, it should in principle vary between 9.5 and 16.4 cm, with $\lambda = 37.3$ cm, the rf wavelength at 805 MHz, the chosen operating rf frequency of the SCL. To optimize the accelerating gradient, and the transit time factor, it would be desirable to manufacture cavities with cells varying in length as the beam accelerates. This may not be economical, and we prefer [2-4] to manufacture rf cavities all with the same cell length. This simplifies the design, and reduces the cost, at the expense of a modest reduction of the transit time factor. Here we assume that the SCL is divided in two subsections each operating at two intermediate values of velocity. A super-conducting MES, from 150 to 300 MeV, has the cavity cell length adjusted to the intermediate value $\beta = 0.525$. The HES, from 300 MeV to 1 GeV, is designed with the intermediate value $\beta = 0.69$.

The layout of the SCL is described in [2, 3]. For more details see also the related paper [5], where a SCL operating in pulsed mode is described. It is made of a sequence of identical periods each consisting of a Warm Insertion for the location of focussing quadrupoles, steering magnets, vacuum pumps, and instrumentation, and of a Cryo-Module including a number of cavities all with the same number of individual cells. One or more cavities are powered by a rf coupler connected to Klystrons, the rf power source.

The general parameters of the SCL are given in Table 1. Cost and other parameters used for our estimate are shown in Table 2. The overall design is summarized in Table 3. The cost items are quoted in year 2000 dollars.

There are four cells in one cavity, each 9.78 and 12.85 cm long, respectively in the MES and HES. There are also six and eight cavities in one cryostat, separated by 32 cm from each other. The cold-to-warm transitions at both ends of the cryostat are 30 cm long. With a 1-meter long warm insertion, this makes a period 5.546 and 7.951 m long, respectively in the MES and HES. To focus the transverse motion of the beam a FODO sequence of quadrupole magnets is assumed, with a quadrupole located every period.

Table 1. General Parameters of the 10-MW, 1-GeV SCL

|  | MES | HES |
|---|---|---|
| Beam Power (CW) | 3.6 MW | 10 MW |
| Beam Current | 12 mA | 10 mA |
| In. Kinetic Energy | 150 MeV | 300 MeV |
| Fin. Kinetic Energy | 300 MeV | 1.0 GeV |
| Frequency | 805 MHz | 805 MHz |
| Protons / Bunch | $3.73 \times 10^8$ | $3.10 \times 10^8$ |
| Temperature | 2.0 °K | 2.0 °K |
| Cells / Cavity | 4 | 4 |
| Cavities / Period | 6 | 8 |
| Cavity Separation | 32 cm | 32 cm |
| Cold-Warm Trans. | 30 cm | 30 cm |
| Cavity Int. Diameter | 10 cm | 10 cm |
| Length of Warm Ins. | 1.00 m | 1.00 m |
| Accel. Gradient | 7.382 MeV/m | 11.234 MeV/m |
| Cavities / Klystron | 6 | 8 |
| rf Couplers / Cavity | 1 | 1 |
| Rf Phase Angle | 30° | 30° |
| Focussing Method | FODO | FODO |
| Phase Advance / cell | 90° | 90° |
| Norm. rms Emitt. | 0.3 π mm mrad | 0.3 π mm mrad |
| Rms Bunch Area | 0.5 π ° MeV | 0.5 π ° MeV |

We have also assumed a cryogenic temperature of the rf cavities of 2 °K. One Klystron groups together all the cavities in one Cryo-Module. The power of one Klystron is divided in 6 or 8 rf couplers, one for each cavity. To avoid saturation of the Klystrons, a 35% rf power contingency has been included above the requirement for the normal mode of operation.

Table 2. Cost ('00 $) and Other Parameters

| AC-to-rf Efficiency | 0.585 | For CW mode |
|---|---|---|
| Cry. Efficiency | 0.004 | At 2.0 °K |
| Electricity Cost | 0.05 | $ / kWh |
| Linac Availability | 75 | % of yearly time |
| Normal Cond. Cost | 150 | k$ / m |
| Superconducting Cost | 500 | k$ / m |
| Tunnel Cost | 100 | k$ / m |
| Cost of Klystron | 1.7 | $ / W of rf Power |
| Cost of Refrig. Plant | 2 | k$ / W @ 2.0 °K |
| Cost of Electrical Distr. | 0.14 | $ / W of AC Power |

Table 3. Summary of the 1.0-GeV SCL Design

| | LES | HES |
|---|---|---|
| Energy: in | 150 MeV | 300 MeV |
| out | 300 MeV | 1 GeV |
| Velocity, β: in | 0.5066 | 0.6526 |
| out | 0.6526 | 0.8750 |
| **Cell Reference $\beta_0$** | **0.525** | **0.69** |
| Cell Length, cm | 9.78 | 12.85 |
| Total No. of Periods | 10 | 18 |
| Length of a period, m | 5.546 | 7.951 |
| FODO ampl. func., $\beta_0$, m | 18.94 | 27.15 |
| **Total Length, m** | **55.46** | **143.13** |
| Coupler rf Power, kW (*) | 40.5 | 67.5 |
| En. Gain/Period, MeV | 15 | 40 |
| Total No. of Klystrons | 10 | 18 |
| Klystron Power, kW (*) | 243 | 540 |
| $Z_0 T_0^2$, ohm/m | 266.7 | 460.6 |
| $Q_0 \times 10^9$ | 5.41 | 7.1 |
| Dissipated Power, kW | 0.89 | 3.02 |
| HOM-Power, kW | 0.048 | 0.079 |
| Cryogenic Power, kW | 1.17 | 3.72 |
| Beam Power, MW | 1.8 | 7.0 |
| Total rf Power, MW (*) | 2.43 | 9.45 |
| AC Power for rf, MW (*) | 4.16 | 16.16 |
| AC Power for Cryo., MW | 0.29 | 0.93 |
| **Total AC Power, MW (*)** | **4.45** | **17.09** |
| **Efficiency, % (*)** | **40** | **41** |
| Capital Cost '00 M$: | | |
| Rf Klystron (*) | 4.13 | 16.07 |
| Electr. Distr. (*) | 0.623 | 2.392 |
| Refrig. Plant | 2.331 | 7.441 |
| Warm Structure | 1.650 | 2.850 |
| Cold Structure | 22.731 | 62.563 |
| Tunnel | 5.546 | 14.313 |
| **Total Cost, '00 M$ (*)** | **37.01** | **105.63** |
| **Oper. Cost, '00 M$/y (*)** | **1.461** | **5.614** |

(*) Including 35% rf power contingency.